\documentclass[iop]{emulateapj} 
\usepackage{amsmath}\usepackage{graphicx}
\usepackage{hyperref}
\usepackage{url}
\usepackage{times}
\usepackage{natbib}
\usepackage{graphicx}
\usepackage{amsmath}
\usepackage{amsfonts}
\usepackage{amssymb}
\usepackage{import}
\usepackage{bibentry}
\usepackage{color}

\newcommand{\edited}[1]{{#1}}

\newcommand{\muram}{MURaM}
\graphicspath{{/}}
\shortauthors{Agrawal et al.}

\begin{document}

\title{Transport of internetwork magnetic flux elements in the solar photosphere}

\author{Piyush Agrawal$^{1,2}$, Mark P. Rast$^{1,2}$, Milan Go\u si\'c$^3$, Luis R. Bellot Rubio$^3$, and Matthias Rempel$^4$}
\affil{$^1$Department of Astrophysical and Planetary Sciences, University of Colorado, Boulder, CO 80309, USA; piyush.agrawal@colorado.edu}
\affil{$^2$Laboratory for Atmospheric and Space Physics, University of Colorado, Boulder, CO 80303, USA; mark.rast@lasp.colorado.edu}
\affil{$^3$Instituto de Astrof\'isica de Andaluc\'ia (CSIC), Apdo. de Correos 3004, E-18080 Granada, Spain}
\affil{$^4$High Altitude Observatory, National Center for Atmospheric Research, Boulder, CO 80307, USA}

\begin{abstract}
The motions of small-scale magnetic flux elements in the solar photosphere can provide some measure of the Lagrangian properties of the convective flow. Measurements of these motions have been critical in estimating the turbulent diffusion coefficient in flux-transport dynamo models and in determining the Alfv\'en wave excitation spectrum for coronal heating models. We examine the motions of internetwork flux elements in a 24 hour long \textit{Hinode}/NFI magnetogram sequence with 90 second cadence, and study both the scaling of their mean squared displacement and the shape of their displacement probability distribution as a function of time. We find that the mean squared displacement scales super-diffusively with a slope of about 1.48.  Super-diffusive scaling has been observed in other studies for temporal increments as small as 5 seconds, increments over which ballistic scaling would be expected. Using high-cadence MURaM simulations, we show that the observed super-diffusive scaling at short temporal increments is a consequence of random changes in the barycenter positions due to flux evolution. We also find that for long temporal increments, beyond granular lifetimes, the observed displacement distribution deviates from that expected for a diffusive process, evolving from Rayleigh to Gaussian. This change in the distribution can be modeled analytically by accounting for supergranular advection along with motions due to granulation. These results complicate the interpretation of magnetic element motions as strictly advective or diffusive on short and long timescales and suggest that measurements of magnetic element motions must be used with caution in turbulent diffusion or wave excitation models. We propose that passive trace motions in measured photospheric flows may yield more robust transport statistics.	
\end{abstract}

\keywords{Sun: photosphere --- Sun: granulation}
\section{Introduction}
\label{sec-intro}

The motions of small-scale magnetic flux elements in the solar photosphere are largely determined by plasma flows. Studying these motions can contribute to our understanding of the Lagrangian dynamics in the radiative boundary layer of the highly turbulent solar convection zone.  This in turn can inform models of coronal heating by Alfv\'en waves, since the spectrum of those waves depends on the `footpoint' motions~\citep[e.g.,][]{cranmer05, ballegooijen14, Vankooten17}.  Additionally, flux-transport models of the solar dynamo rely on cross equatorial reconnection of the opposite polarity field along with poleward transport of the residual to reverse the sign of the global field every half cycle period~\citep[e.g.,][]{babcock55, wang89-cycle21, dikpati07, jiang14}. These processes are often modeled as due to the combined action of meridional flow and supergranular diffusion.  

For low molecular diffusivities, turbulent transport in the continuum approximation can be described in terms of Lagrangian parcel motions~\citep[e.g.,][]{toschi09}. 
This approach faces some challenges in the context of solar magnetic flux elements because the magnetic field motion is not strictly passive but back-reacts on the flow, the two-dimensional motions observed in the photosphere represent some unknown average of the flow over the range of depths to which the field extends, and the elements themselves have finite lifetimes. Of these, the first effect may be small because the ratio of plasma to magnetic energy density in the quiet-sun photosphere is large, the second may contribute to the observed field strength dependence of the element motions \citep[e.g.,][]{hagenaar99, Yang15}, and the last may make the interpretation of element transport as a diffusive process more challenging~\citep[e.g.,][]{yuste13}, though this issue has not yet been studied in the solar context. While it is important that these difficulties be examined in future work, here, as a first approximation and in common with most previous studies, we treat the magnetic flux elements as Lagrangian tracers to understand the implications of their motions under that assumption. 

In that context, transport is usually characterized by how the mean squared displacement of the flux-weighted barycenter of magnetic elements scales with time \citep[e.g.,][]{lawrence93}, $\langle r^2 \rangle\propto t^\gamma$, where $r$ is the Lagrangian displacement of each element over a temporal increment $t$, and $\gamma$ is the inferred scaling exponent. 
The temporal increment, or time interval, refers to the time elapsed since any moment along the trajectory of a flux element, not just the time since its emergence. Thus, for a given temporal increment, multiple displacement measurements are possible when the trajectory spans a length of time longer than the temporal increment being considered. 
For very short time intervals, below the Lagrangian integral time (the autocorrelation time of the velocity along a parcel trajectory), the motion of any single flux element is expected to be highly correlated and the mean squared displacement of all the elements should scale ballistically ($\gamma = 2$).  Subsequently, as the Lagrangian motions de-correlate, both because the trajectories spatially sample a wider range of the flow, which has a finite spatial correlation length, and because the flow itself temporally de-correlates, an intermediate value of $1 < \gamma < 2$ is expected.  This scaling is sometimes referred to as super-diffusive.  Finally, over intervals longer than the Eulerian integral time (the autocorrelation time of the velocity at a fixed point), the Lagrangian motions are expected to be fully de-correlated, and the displacements of the flux elements should display a random walk with diffusive scaling $\gamma=1$.  Once measured, $\gamma$ can be used to calculate an effective diffusion coefficient \citep{abramenko11}, but, as we shall see, the physical interpretation of the process as diffusion may be problematic. 

\edited{Table \ref{table-drdt} summarizes the measured values of $\gamma$ from recent studies, where the studies are limited to those of internetwork elements, the focus of this paper. The values of $\gamma$ vary depending on the data set examined and the feature tracking algorithm employed, but what is peculiar is that the scaling, even at very short temporal increments, is super-diffusive~\citep[e.g.,][]{abramenko11, chitta12, jafarzadeh14}, even though one would expect these increments to be significantly shorter than both the Eulerian and Lagrangian integral times. }

\edited{The scaling of the mean-squared displacement with time provides only limited information about the underlying flows which guide the motion of the flux elements. The probability distribution of the Lagrangian displacements is more sensitive to the flow properties and can capture some of the effects of turbulent intermittency~\citep{rast16}.  For very short temporal increments, over which the mean squared displacement of the flux elements should scale ballistically, the probability distribution of the displacements should reflect the underlying Lagrangian velocity distribution.  As the motions of the flux elements de-correlate and the scaling becomes diffusive, the probability distribution should, for a two-dimensional motion, approach a Rayleigh distribution (the distribution would be Maxwellian for three-dimensional motions). We find that displacement distributions of internetwork magnetic elements do not behave as expected.} 

\edited{In~\S\ref{sec-model} we examine the displacement probability distribution of magnetic elements as a function of time. While the distribution approaches Rayleigh as the trajectories de-correlate over granular lifetimes, at longest intervals ($t\gtrsim2$ hours), it surprisingly becomes Gaussian. With the help of a simple correlated random walk model with drift motion we demonstrate that this is likely due to the presence of an underlying large-scale supergranular flow which dominates granular motions over long timescales.  We expect that transport by supergranular motions on timescales long compared to their lifetimes would be similarly affected by the underlying meridional flow. The multiscale and intermittent nature of the flows leads to flux transport that cannot be described as either strictly diffusive or purely advective, scaling is thus subdominant~\citep{rast11}.  In~\S\ref{sec-muram} we revisit the observed discrepancy in scaling at shortest times and, with the help of a radiative magnetohydrodynamic simulation of solar granulation, show that the observed super-diffusive scaling is likely an artifact of random changes in the barycenter positions of the magnetic elements, induced by rapid changes in their flux content or configuration.}

\begin{table}[t]
	\centering
	\caption{Recent work on the transport of internetwork magnetic flux elements.}  
	\label{table-drdt}
	\begin{tabular}{llll} 
		\hline
		Author                 	& Instrument   	& Cadence (s)  	& $\gamma^a$ \\ 
		\hline
		\cite{abramenko11}$^b$      & BBSO/NST      & 10       	& 1.48-1.67   \\
		\cite{chitta12}$^b$         & SST/CRISP     & 5        	& 1.59 		  \\
		\cite{giannattasio14a}$^c$  &  \textit{Hinode}/SOT   & 90 		& 1.55      	\\ 
		\cite{giannattasio14b}$^c$  & \textit{Hinode}/SOT 	& 90 		& 1.44      	\\ 
\edited{\cite{jafarzadeh14}}$^b$        & Sunrise/SuFI         & 3-12      & 1.69          \\
		\cite{manso11}$^c$          & \textit{Hinode}/SOT   	& 28 		& 0.96, 1.70 	\\
		
		\\
		This work$^c$               & \textit{Hinode}/NFI    & 90 		& 1.48 		  	\\
		\hline
	\end{tabular} \\
	\raggedright{$^{a}$Scaling exponent $\gamma$ computed using $^b$magnetic bright points and $^c$magnetic flux elements.}
\end{table}

\section{Observational data and results}
\label{sec-observations}

To investigate the transport of small-scale internetwork magnetic flux elements, we analyzed observations obtained with the Narrowband Filter Imager \cite[NFI;][]{hinodesot08} on-board the Hinode spacecraft~\citep{hinodemission07}. The data sequence is a part of the Hinode Operation Plan 151 (HOP 151). The measurements are well suited to study the quiet-sun magnetic fields on temporal scales from minutes to days because of their high spatial and temporal resolution, flux sensitivity, and long duration.

The data are an uninterrupted time sequence of magnetograms, about 24 hours long, starting at 08:32:00~UT on 23 November 2010, taken with 90 second cadence.  They cover a field of view of about $41\times46{\rm \ Mm}^2$, with $0".16$ ($116$ km) pixel size, and were constructed using Stoke I and V measurements $\pm160$ m\AA\, from the 589.6 nm \ion{Na}{1} D1 line core center.  Post processing of the data and removal of the $p$-mode signal by the application of a subsonic filter~\citep{Title89, Straus92} yields magnetograms with a noise level of 4 Gauss, an estimate based on the standard-deviation of pixels with no clear magnetic signal. More details regarding data processing and calibration can be found in \cite{gosic14}. 

\begin{figure}[t]
           \centering
            \includegraphics[width=8cm, keepaspectratio]{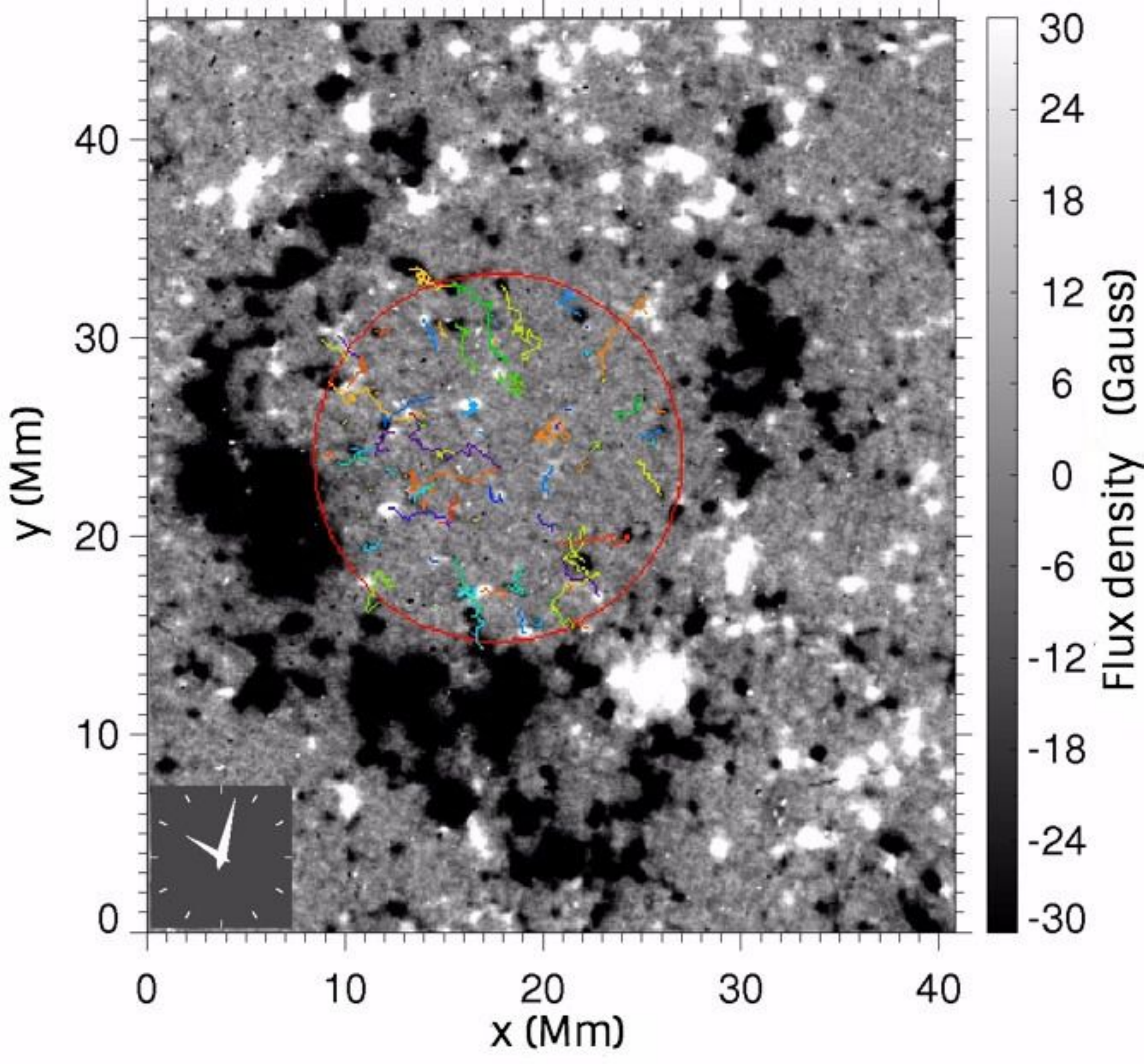} 
          \caption{Sample \textit{Hinode}/NFI magnetogram saturated at $\pm30$ Gauss. The red circle marks the boundary of the internetwork region. Colored curves are the trajectories of individual flux elements.}             		
          \label{fig-luis-mag}          
\end{figure}

The magnetogram sequence employed in this study samples the quiet sun at disk center, capturing the spatio-temporal evolution of a single supergranule and its surroundings.  A single frame from the time series is shown in Figure~\ref{fig-luis-mag}. The red circle, with radius 9.3 Mm (corresponding to 0.8 times that of the supergranular cell~\citep{gosic16}), outlines the internetwork region within which flux elements were identified and tracked.  The multi-color curves indicate the trajectories of some representative flux elements.

\subsection{Tracking Algorithm}
\label{subsec-tracking}

We used a semi-automatic procedure to determine the trajectories of the magnetic flux elements. Identification and tracking were carried out automatically, but the results were revised manually at each time-step to verify the performance of the code. In cases of element misidentification or interaction (merging, cancellation or fragmentation), the code output was corrected and the tracking continued from that step until precise trajectories were derived. The feature identification algorithm employed the clumping method of~\cite{Parnell09}, with a minimum unsigned flux density of 12 Gauss (3 times the noise level) and a minimum element area of 4 pixels. This yielded elements with a mean unsigned flux density of about 23 Gauss. To identify individual elements during interactions we used the downhill method of~\cite{welsch03}, which allowed us to maintain their identification over long periods of time. Our tracking approach thus overcomes some of the difficulties faced by standard algorithms during element interactions~\citep[e.g.,][]{gosic14, gosic16}. The tracking was repeated twice to check for consistency, and nearly identical results were obtained each time. Further details can be found in~\cite{gosic_master}).

For the analysis that follows, we restrict identification to elements whose trajectories begin within the internetwork region defined by the red circle in Figure~\ref{fig-luis-mag}.  Further,  element lifetimes ranged from 1.5 minutes to 5.5 hours, and we retained for analysis only those with lifetimes $\geq$ 4.5 minutes. This yields a total of 6463 unique magnetic flux element trajectories over the 958 magnetogram sequence.  The change in flux-weighted barycenter positions of these elements as a function of time forms the basis for the displacement statistics analyzed below.   

\subsection{Mean Squared Displacement}
\label{sec-msd}
\begin{figure}[t]
     \includegraphics[width=8cm, keepaspectratio]{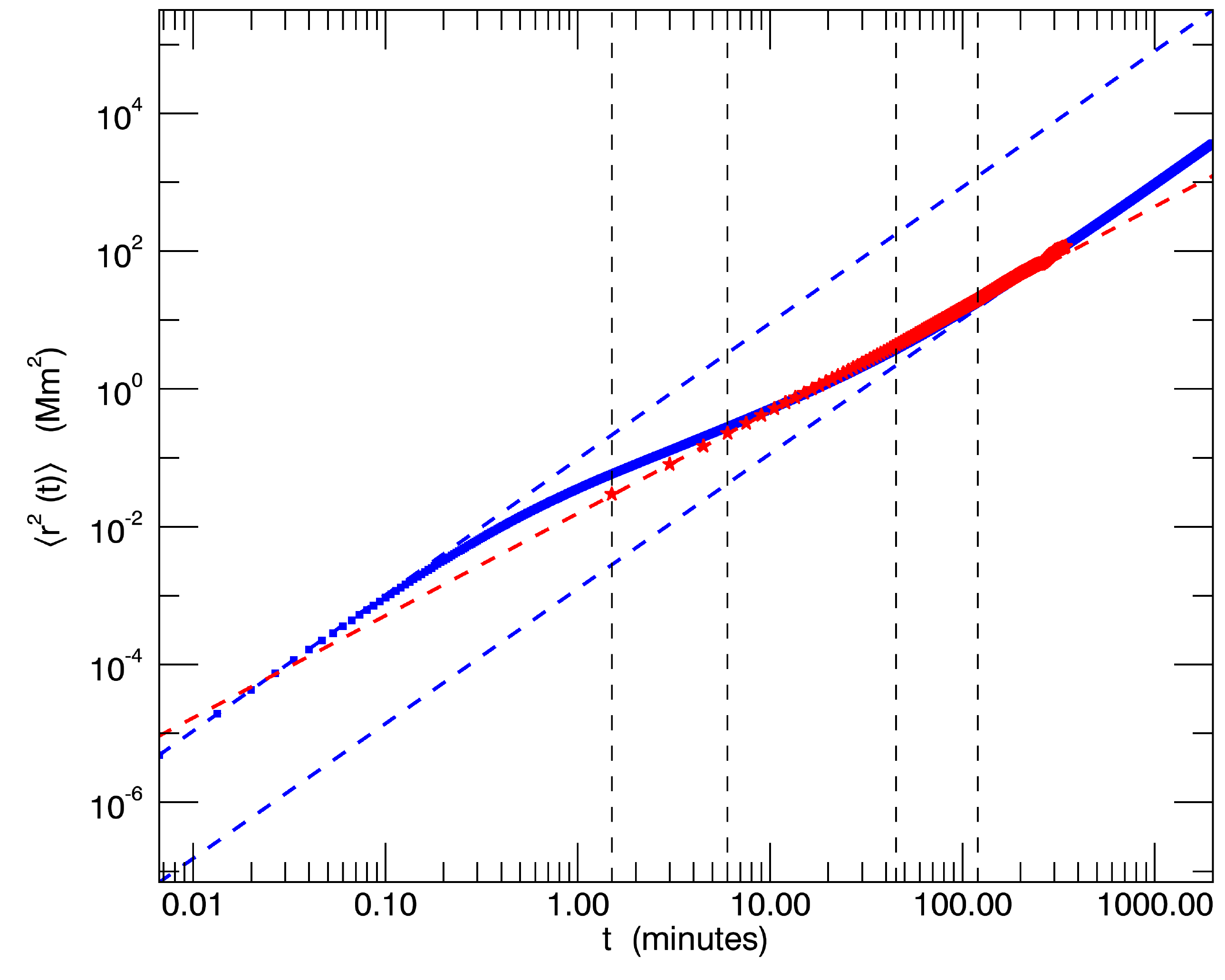} 
     \caption{Mean squared displacement as a function of temporal increment for the magnetic flux elements in \textit{Hinode}/NFI data ({\it red stars}) and for the random walk model ({\it blue squares}) of~\S\ref{subsec-model-dist}.  The best-fit slope for the \textit{Hinode}/NFI data is 1.48 ({\it red dashed line}), computed from fitting the data below 2 hours. The scaling for the model is 1.97 and 1.96 ({\it blue dashed lines}) at the shortest and the longest temporal increments, respectively.  The vertical \textit{black dashed lines} correspond to the increments for which the displacement probability distributions are shown in Figure~\ref{fig-pdfs}. For both observations and the model, the variance of the squared displacement at each temporal increment does not exceed the size of the plotting symbol.}
     \label{fig-drdt}
    \end{figure}

The mean squared displacement of the flux elements is plotted as a function of temporal increment in Figure~\ref{fig-drdt} (\textit{red stars}). 
It approximates a power law with a super-diffusive scaling exponent, $\gamma=1.48$.  This is consistent with previous measurements (Table~\ref{table-drdt}), the value falling in-between ballistic and diffusive values. There is some deviation from the power-law behavior for time intervals greater than about 2 hours which exceeds the displacement variance.  In~\S\ref{subsec-model-dist} we suggest that this deviation at long times reflects advection by the underlying supergranular flow, for which there is strong evidence in the displacement distributions. 

\subsection{Displacement Probability Distributions}
\label{sec-pdf}

\begin{figure*}[tbh]
	\includegraphics[width=18cm, keepaspectratio]{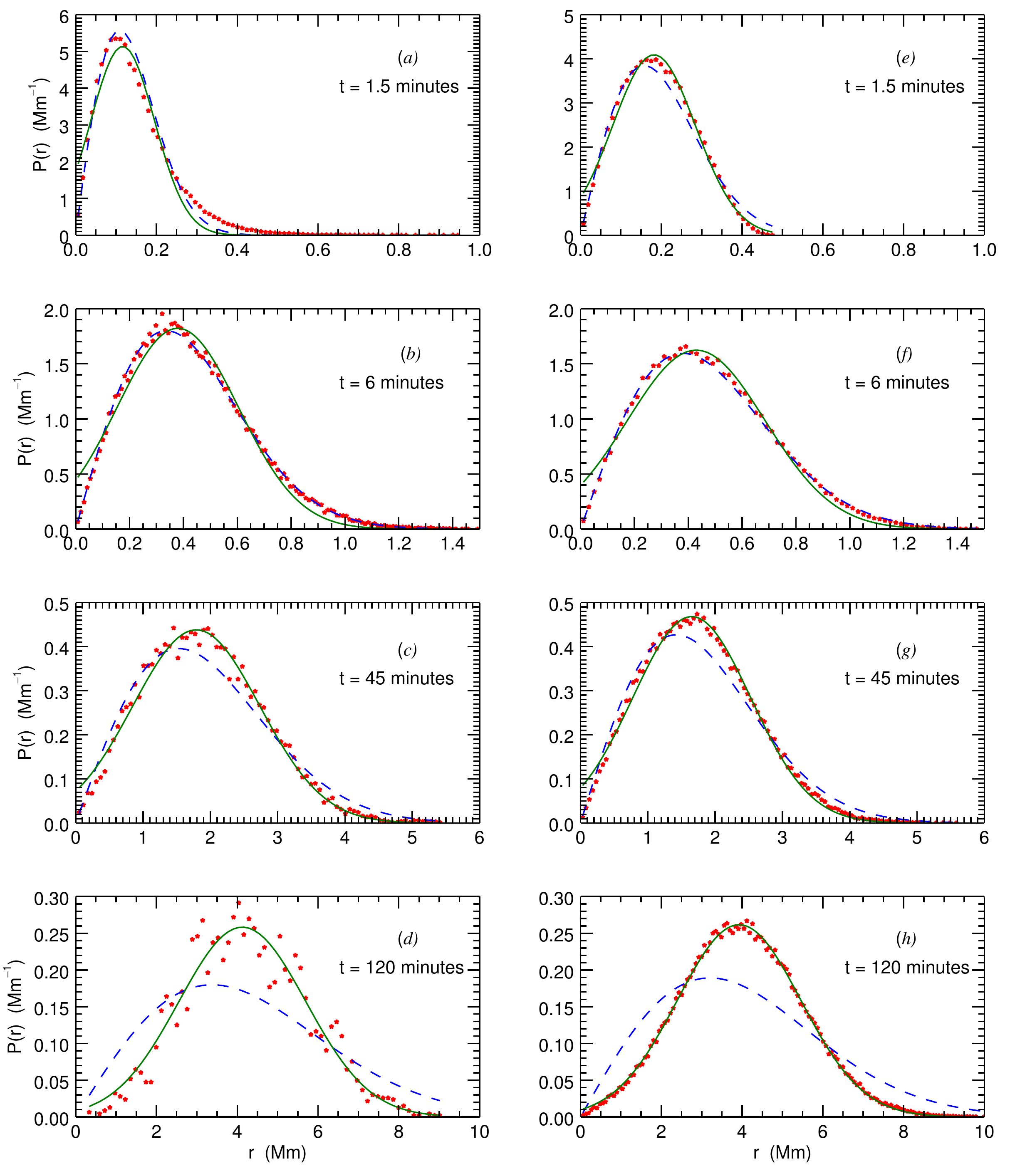} %
	\caption{Displacement probability distributions for \textit{Hinode}/NFI data (left column) and for the random walk model (right column) of~\S\ref{subsec-model-dist}, at different temporal increments, are shown here. The \textit{red stars} are the distribution values, the \textit{blue (dashed)} and the \textit{green (solid)} curves refer to the best-fit Rayleigh and Gaussian curves, respectively.  Note that only $\sim2\%$ of the flux elements survive to contribute to the \textit{Hinode}/NFI data distribution at 120 minutes, and since the bin size is held constant at about 0.13 Mm, the histogram is noisy.}
	\label{fig-pdfs}
\end{figure*}  

While the mean squared displacement of the flux elements closely follows a power-law for time intervals below 2 hours (Figure~\ref{fig-drdt}), the underlying probability distribution of the displacements evolves over this range of increments. The left column of Figure~\ref{fig-pdfs} shows the observed distributions for four different temporal increments, $t=1.5, \ 6, \ 45$ and $120$ minutes, along with the best-fit Rayleigh and Gaussian functions ({\it blue (dashed)} and {\it green (solid)} curves, respectively).  
For short increments, $t=1.5$ minutes (Figure~\ref{fig-pdfs}$a$), the distribution is neither Rayleigh nor Gaussian, but shows an elevated probability for large displacements. In~\S\ref{sec-muram} we argue that this is likely due to the apparent motion of the magnetic element barycenters when they are subject to flux evolution and element identification uncertainties.  For intermediate increments, $t=6$ minutes (Figure~\ref{fig-pdfs}$b$), the distribution approximates a Rayleigh distribution because the motions have largely de-correlated.  Somewhat surprisingly, for longer temporal increments the distribution does not remain Rayleigh ($t=45$ minutes, Figure~\ref{fig-pdfs}$c$) but instead becomes nearly Gaussian ($t=120$ minutes, Figure~\ref{fig-pdfs}$d$).  We show below that a larger scale drift component (likely originating with supergranulation in the observations) superimposed on the granular motion can explain both this change and the deviation from power-law scaling at long intervals.   


\section{Model: Correlated Random Walk with Drift}
\label{sec-model}

\subsection{Model Definition} 
\label{sec-model-const}
       
Under the assumption that magnetic flux elements are advected passively by the underlying plasma flow, we model the effects of granular and supergranular flows on magnetic element motions as random walk and drift contributions to the motion of Lagrangian `walkers'. The effect of granular flow is modeled as a correlated random walk, a random walk with an imposed correlation in the step direction.  The random walk steps are of size $v_{\rm g}\,\Delta t$, where both $v_{\rm g}$ and the step interval $\Delta t$ are prescribed constants. The direction of travel of a walker is taken to be $\theta+\Delta\theta$, where $\theta$ is the direction of motion during the previous step and $\Delta\theta$, computed anew for each time-step, is a uniformly sampled random variable between $-\pi/C$ and $\pi/C$.  The parameter $C$ constrains the step direction (induces memory) and thus controls the time it takes for the motions to de-correlate. For  $C = 1$, $\Delta\theta\in[-\pi , \pi)$ and the motion is delta-correlated (step direction de-correlates after one time-step). If $C$ is large, the change in direction of the walker ($\Delta\theta$) at each time-step is small, and the motion stays correlated for a longer time. When interpreting observations in light of this model, the parameter $C$ is chosen so that the random walk component of the motion de-correlates over a granular lifetime. 

To keep the number of free model parameters to a minimum, the supergranular contribution to the magnetic element motions is modeled as a uniform constant drift in a single direction. This approximation is reasonable as long as the elements being tracked are internetwork elements (as is the case for our data) and the supergranular flow is approximately spatially uniform and steady over any individual granulation induced random walk trajectory.  This is true if the supergranular motions are directed approximately radially away from their centers and, over the lifetime of an individual flux element, the granular motions do not induce a significant deviation from that radial direction. We can determine if our model is self-consistent by checking if these conditions are met for the parameters that approximate granular and supergranular contributions.  For these (see \S\ref{subsec-model-dist}), we find that the average deviation of a random walk trajectory from the direction of drift is about $\pm 2.3^\circ$ over the time it takes for the drift component to carry the walkers 30 Mm (the characteristic supergranular length scale, \cite{LR-rieutord10}). Thus, motion due to granulation approximately samples only a small region about the mean supergranular drift direction, though radial gradients in the supergranular flow may pose difficulties for modeling very long-lived element displacements that traverse the full supergranular extent.  The same drift can be applied to each random walk realization, even if in reality the supergranular flow carries each magnetic element in a radially different direction, because the granular component carries the individual walkers in all directions, and only the distance traveled, not the direction of travel, is of interest to the displacement statistics.  In practice, when constructing the two-dimensional correlated random walk model, we choose the drift to act in the positive $x$-direction and vectorially add $v_{\rm sg}\,\Delta t$ to the correlated random walk component $v_{\rm g}\,\Delta t$ at each time-step, where $v_{\rm sg}$ is the drift velocity.
       
\subsection{Model Displacement Probability Distributions} 
\label{subsec-model-dist}

To qualitatively compare model results with observations, we take $C=7$, $v_{\rm g}=5.5$ km/s and $v_{\rm sg}=0.5$ km/s. When scaled with $\Delta t=0.4$ seconds, the de-correlation time of random walk component is on average about 6 minutes. These values have not been fine tuned, but are reasonably representative of granular and supergranular horizontal flow velocities and typical granule lifetimes \citep[e.g.,][]{rast03, LR-nordlund09,LR-rieutord10}.
Using these parameters, we compute the trajectories of the walkers and determine their displacements as a function of temporal increment. The resulting displacement distributions (Figure~\ref{fig-pdfs}e-h) qualitatively agree with the observations for all but the shortest increments.  
As mentioned previously, at these shortest times the observed distribution reflects the underlying Lagrangian velocity distribution and any pathologies associated with tracking the barycenter of magnetic elements. 
The later issue does not enter this simplified correlated random walk model, and we discuss its role in the observations in more detail in~\S\ref{sec-muram}.      
It is worth noting here, that even in the absence of the barycenter complications, the observed distribution at early times is unlikely to be captured by this simplified model since
unlike for the observations, the Lagrangian velocity distribution in the model is a delta function.  What is important is that this simple model captures the evolution of the distribution at longer times.  As is the case for magnetic elements in the \textit{Hinode}/NFI data, the distribution first becomes Rayleigh, as the motions de-correlate over temporal increments longer than the Lagrangian autocorrelation time, and then, for still longer increments, approaches an offset Gaussian, as the slow drift comes to dominate the displacement.   

\begin{figure*}[t]
	\centering
	\includegraphics[width=18cm, height=6cm, keepaspectratio]{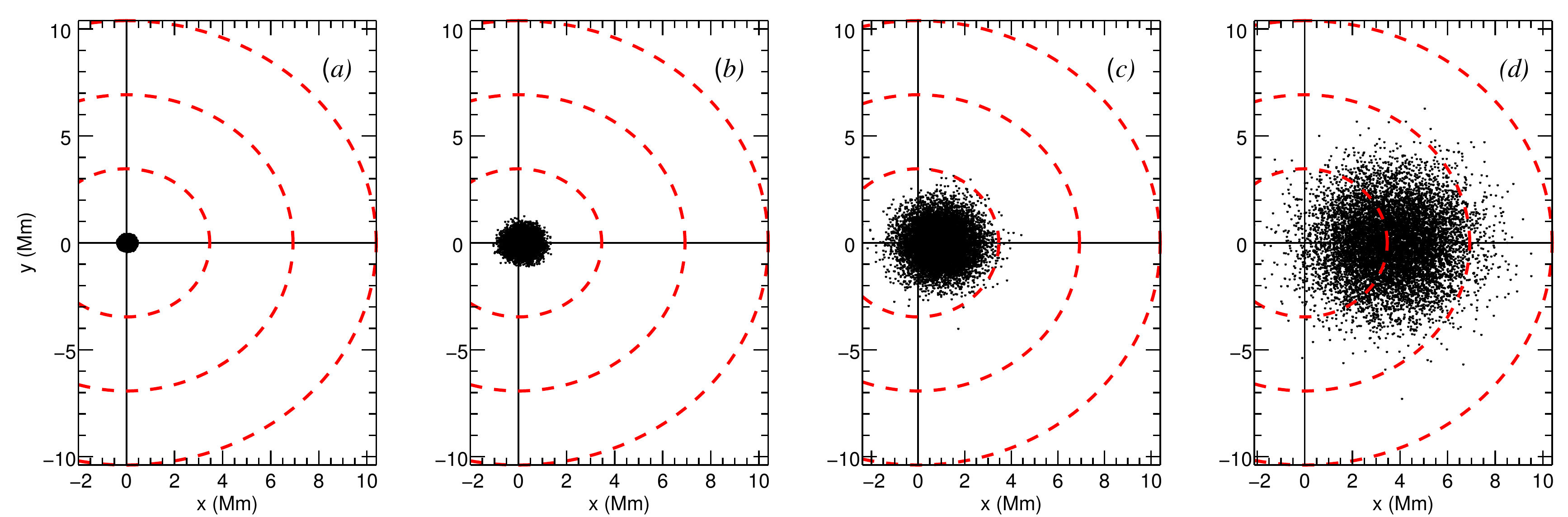} 
	\caption{Position of the model walkers (\textit{black points}) at t = (a) 1.5, (b) 6, (c) 45 and (d) 120 minutes. The \textit{Red (dashed)} concentric circles are contours of constant radius and emphasize the drift of walkers along positive \textit{x}-direction, away from the origin. \textit{Black solid} lines mark the \textit{x} and \textit{y}-axis.}
	\label{fig-walker-position} 
\end{figure*} 

This dependence of the shape of the probability distribution function on temporal increment can be expressed analytically.  For time intervals long compared to the granular correlation time, the walker motions combine a random walk with a drift. The former yields a two-dimensional Gaussian spread of the walkers positions about the origin while the latter advects the origin downstream.  By bivariate transformation of random variables~\citep[e.g.,][and this paper's Appendix~\ref{appendixa}]{casella2002statistical, hogg2006probability, Rast09}, for spatial offsets in $x$ and $y$ which are Gaussianly distributed with equal variances $\sigma$ about $x_0$ and $y_0$, the radial distance $r$ from a fixed origin at $(0,0)$ is 
distributed as 
\begin{equation}
P(r)= \frac{r}{\sigma^2} \ e^{-(r-r_0)^2/2\sigma^2}\ e^{-r_0r/\sigma^2}\ I_0(r_0r/\sigma^2)\ ,
\label{eqn-pdf1}
\end{equation}
where $r_0^2={x_0^2+y_0^2}$ and $I_0$ is the lowest order modified Bessel function of the first kind~\citep{book-abramowitz72}.  In our model of the Lagrangian displacement of the walkers, both $r_0$ and $\sigma$ increase with temporal increment due to the drift and random walk components of the motion, respectively.  For small values of $r_0$ compared to $\sigma$, the product of the last two terms in Equation~(\ref{eqn-pdf1}) approaches one and the distribution becomes Rayleigh, as expected for a two-dimensional random walk centered at origin.  For large $r_0$ and small $\sigma$, $I_0(r_0r/\sigma^2)\sim\exp(r_0r/\sigma^2)/\sqrt{2\pi r_0r/\sigma^2}$~\citep{book-abramowitz72}, and the distribution becomes 
\begin{equation}
P(r)\sim\sqrt{\frac{(r/r_0)}{2 \pi \sigma^2}} e^{-(r-r_0)^2/2 \sigma^2}\ .
\label{eqn-pdf2}
\end{equation}
Thus for large $r_0$ and small $\sigma$, the distribution is nearly Gaussian around $r=r_0$, with only slight distortion by the $r/r_0$ pre-factor to ensure that $P(r)=0$ at $r=0$. Note that for a random walk without drift, $x_0$ and $y_0$ are typically taken to be zero and the resulting distribution is Rayleigh for all times, but for our case with finite drift contribution $r_0=x_0=v_{\rm sg}t$, the distribution evolves from Rayleigh to Gaussian at long times (Figure~\ref{fig-pdfs}e-h). 

\edited{Figure~\ref{fig-walker-position}a-d displays the positions of the random walkers (\textit{black points}) in our correlated random walk model at times corresponding to the displacement distributions shown in Figure~\ref{fig-pdfs}e-h.  Over very short time intervals, the trajectories are radially ballistic, directed away from the origin.  Over longer intervals, walkers' positions are distributed as a two-dimensional Gaussian `cloud' about the drift position $r_0 = v_{\rm sg}t$, and the displacement distribution about that position is Rayleigh. 
As $r_0$ is small compared to $\sigma$, the peak of the Gaussian cloud is still close to the origin, and the displacements, when computed from the origin, are approximately Rayleigh distributed (see Figure~\ref{fig-walker-position}b and its corresponding displacement distribution in Figure~\ref{fig-pdfs}f). Since the standard-deviation $\sigma$ of the random walk component increases as $t^{1/2}$, while $r_0$ due to the drift increases as $t$, after sufficiently long times the Gaussian cloud drifts away from the origin and the shape of the displacement distribution, as computed from the origin, changes from Rayleigh to nearly Gaussian (see Figure~\ref{fig-walker-position}d and its corresponding displacement distribution in Figure~\ref{fig-pdfs}h). That change is apparent with increasing time in both the model distributions (right column of Figure~\ref{fig-pdfs}) and those derived from the observations (left column of Figure~\ref{fig-pdfs}).}  

The change is also reflected in the mean squared displacement vs. time curve in Figure~\ref{fig-drdt}, in which the curve for walkers ({\it blue squares}) is plotted along with measurements from \textit{Hinode} data. For the shortest temporal increments, the walker motions are highly correlated and the scaling is ballistic.  For longer increments the walker motions de-correlate and the scaling becomes super-diffusive, flattening towards what would, in the non-drifting case, become the diffusive value. With drift, this is interrupted, and the scaling reverts to ballistic as drift contribution to the displacement comes to dominate the motion.  Power-law fits to the model data at the shortest and longest times have indices of 1.97 and 1.96, respectively.  This reversion to ballistic scaling occurs for temporal increments longer than $(v_g/v_{sg})^2 * \tau_{g}$ if the drift is steady, where $\tau_{g}$ is a typical granule correlation time.  
If the correlation time of the random component and the timescale on which drift dominates the motions are sufficiently well separated, then a diffusive scaling between the two ballistic regimes can be achieved. This is not the case here because the ratio of the granular random walk velocity to the supergranular drift velocity is insufficiently large.  For temporal increments greater than 2 hours, the \textit{Hinode} data may show the beginning of the change in scaling.  Unfortunately, the flux elements do not have sufficiently long lifetimes to recover the full reversion to a ballistic scaling.

\section{Flux elements and Passive tracers in MURaM simulations}
\label{sec-muram}

As noted previously, the model of~\S\ref{sec-model}, based on a correlated random walk with drift, does not capture the observed mean squared displacement scaling for magnetic elements at shortest time intervals.  The observed scaling at shortest time intervals is super-diffusive, rather than ballistic, as in the simplified model. This is true for element trajectories in our \textit{Hinode} observations and for trajectories determined using data with cadences as high as $5$ seconds~\citep[][Table~\ref{table-drdt}]{chitta12}.  One expects these time intervals to be well below the Lagrangian and Eulerian integral times, and under the assumption that flux elements are advected passively, the element displacements should thus scale ballistically. 

To uncover the origin of this discrepancy, we analyzed a small-scale dynamo quiet-sun simulation using a modified version of the \muram\ radiative magnetohydrodynamics code~\citep{vogler05}. The simulation is similar to the run \textit{O16b} described in \cite{rempel14}, but with the vertical domain size extended to 1.7 Mm above the photosphere and the radiative transfer computed using four opacity bins.  The overall domain size is about $6\times6\times4$ Mm$^3$, with a uniform grid spacing of 16 km. The simulation has no imposed mean magnetic field, but the mixed field is generated and maintained by a small-scale dynamo. The average unsigned vertical magnetic flux density at optical depth unity has a value of about 80 Gauss, which is quite representative of the quiet-sun magnetism \citep[see e.g.,][]{Suarez07, Danilovic16}.  The simulation spans about one hour with 2.0625 second cadence (1801 snapshots).  We note that the simulation captures granular flows only, not those at larger scales, so the contribution of supergranulation, critical to transport over longer time intervals (\S\ref{subsec-model-dist} above), cannot be examined.  In this section, we focus on time intervals below 100 seconds for which we expect the scaling to be independent of supergranular motions, and are particularly interested in the shortest temporal increments for which ballistic scaling is expected but not seen in the observations.

We identified and tracked flux elements in the simulations using vertical magnetic flux density images at optical depth unity, employing a minimum unsigned flux density threshold of three times the standard-deviation above the mean.  The threshold was computed independently for each image.  At the native resolution of the simulation, this corresponds to pixels with unsigned flux density in excess of about $470$ Gauss. We note that the threshold values differ for the simulation and the observational data (\S\ref{subsec-tracking}) due to differing image resolution and differences in the optical depth surface on which the elements are tracked.  Only magnetic elements with areas greater than 20 pixels and lifetimes in excess of 10 time-steps (about 20 seconds) were included in the study of the displacement statistics in the simulations. We discarded elements that split or merged over the 100 second interval in order to minimize uncertainty in the displacement measurement, which was based on changes in the flux-weighted barycenter positions.

\begin{figure}[t]
	\centering
	\includegraphics[width=8cm]{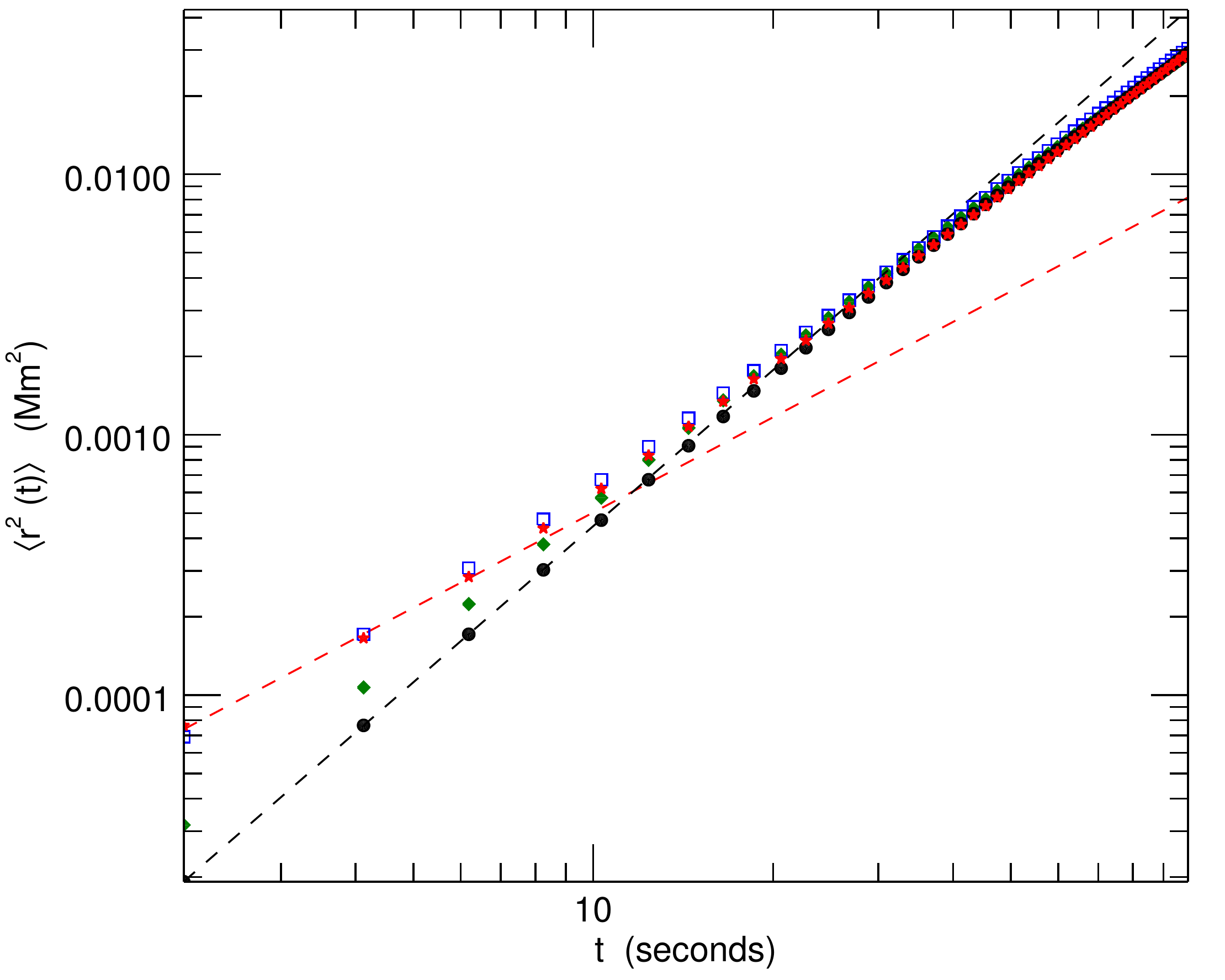} 
	\caption{Mean squared displacement vs. temporal increment for flux elements in \muram\ simulations tracked at native resolution (\textit{red stars}), tracked at degraded 116 km resolution (\textit{green diamonds}), passive tracers (\textit{black dots}) and the model presented in~\S\ref{sec-muram}  (\textit{blue squares}). The best-fit slope at the shortest temporal increments for flux elements tracked at native resolution is 1.21 (\textit{red dashed} line) and that for passive tracers is 1.99 (\textit{black dashed} line). \edited{Note that the plots for  \textit{black dots} and \textit{green diamonds} are shifted vertically upward to better compare their shape with other curves at long increments}. The variance of the mean squared displacements for all the curves do not exceed the size of the plotting symbol.}
	\label{fig-muram-drdt}
\end{figure}

Figure~\ref{fig-muram-drdt} plots the mean squared displacement as a function of temporal increment for the magnetic elements tracked in the \muram\ simulation (\textit{red stars}). The curve approximates a power-law with slope 1.69, similar to the scaling found using observations of magnetic bright points with similar cadence~\citep[e.g.,][Table~\ref{table-drdt}]{chitta12, abramenko11}.  Moreover, the shape of the curve (deviating from a power-law in detail) is similar to that seen in observations~(compare with Figure 8 in~\cite{chitta12} and Figure 5 in~\cite{abramenko11}).  Both the observations and the simulations show a shallower slope for shortest temporal increments, with the scaling exponent in the \muram\ simulation at shortest increments having a value of 1.21.   

\begin{figure}[t] 
	\centering
	\includegraphics[width = 8cm, keepaspectratio]{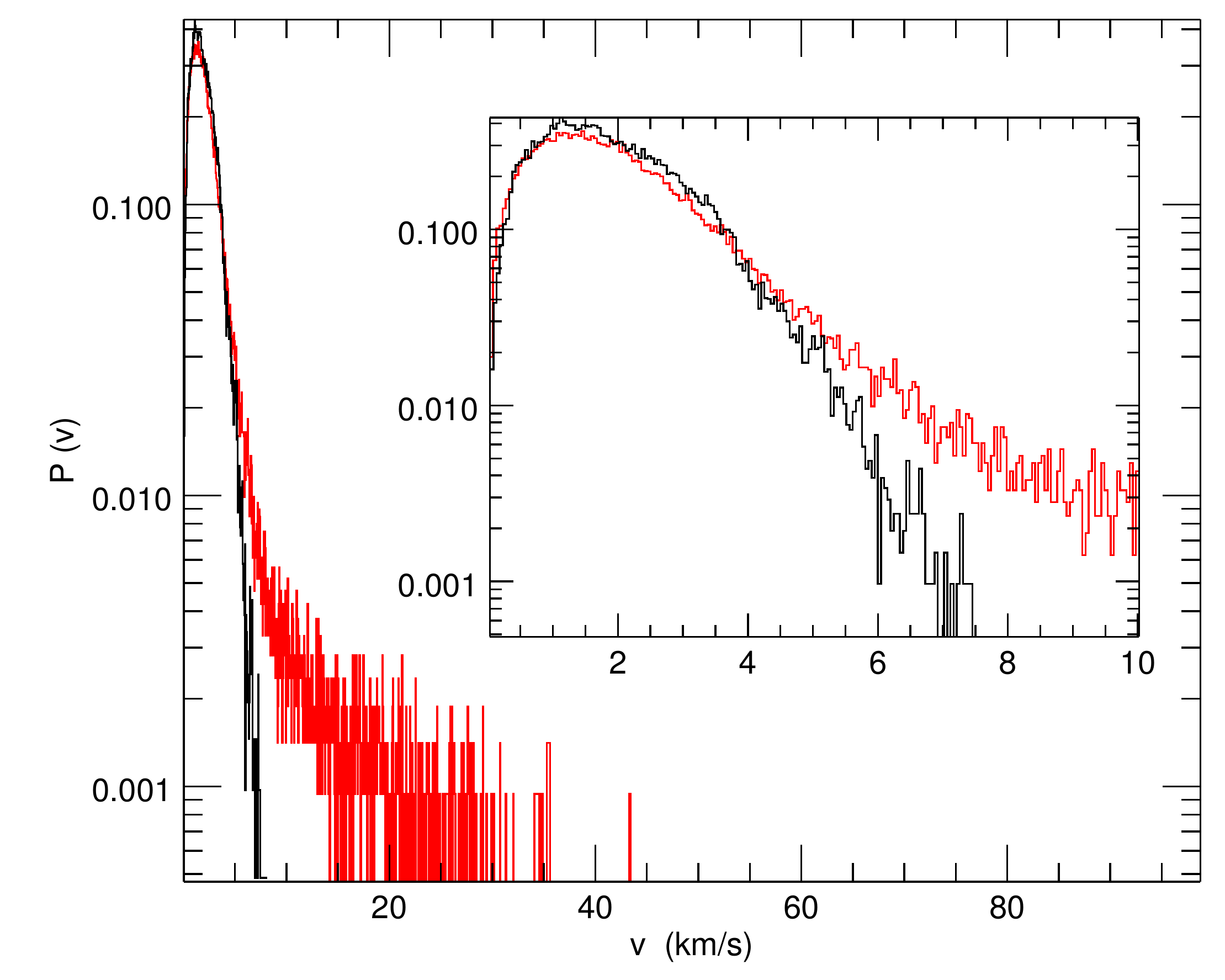} 
	\caption{Lagrangian velocity distribution of the flux elements (\textit{red}) and the passive tracers (\textit{black}) tracked in the \muram\ simulations. Plot in the inset shows the distribution for velocities less than 10 km/s.}
	\label{fig-muram-pdf-vel}
\end{figure}

Not only is the super-diffusive scaling at such short increments surprising in itself, but the increase in the scaling exponent with increasing increment in the absence of any large-scale flow (or in situations where the contribution from the large-scale flow is negligible, as is the case for observations at these increments) is also hard to understand.  
Without a large-scale flow component, the scaling exponent should decrease with increasing increment as the motions de-correlate with time. 
Moreover, oddities are found in the Lagrangian velocity distribution of the magnetic elements in the simulation. The \textit{red} curve in Figure~\ref{fig-muram-pdf-vel} plots the Lagrangian velocity distribution determined from the flux element barycenter displacements after one time-step.  It shows unrealistically high velocity values, well in excess of the photospheric sound speed ($\sim7$ km/s, e.g., \cite{LR-nordlund09}). To check that this is not an artifact of our tracking algorithm, we analyzed magnetic bright point displacements in the photospheric continuum intensity images of the \muram\ solution. The tracking was done by Samuel Van Kooten using the algorithm described in \cite{Vankooten17}.  That algorithm is independent of the one we employed, yet we found the velocity and displacement statistics in agreement with those of the magnetic flux elements presented in this work.

\begin{figure}[t] 
	\centering
	\includegraphics[width = 8cm, keepaspectratio]{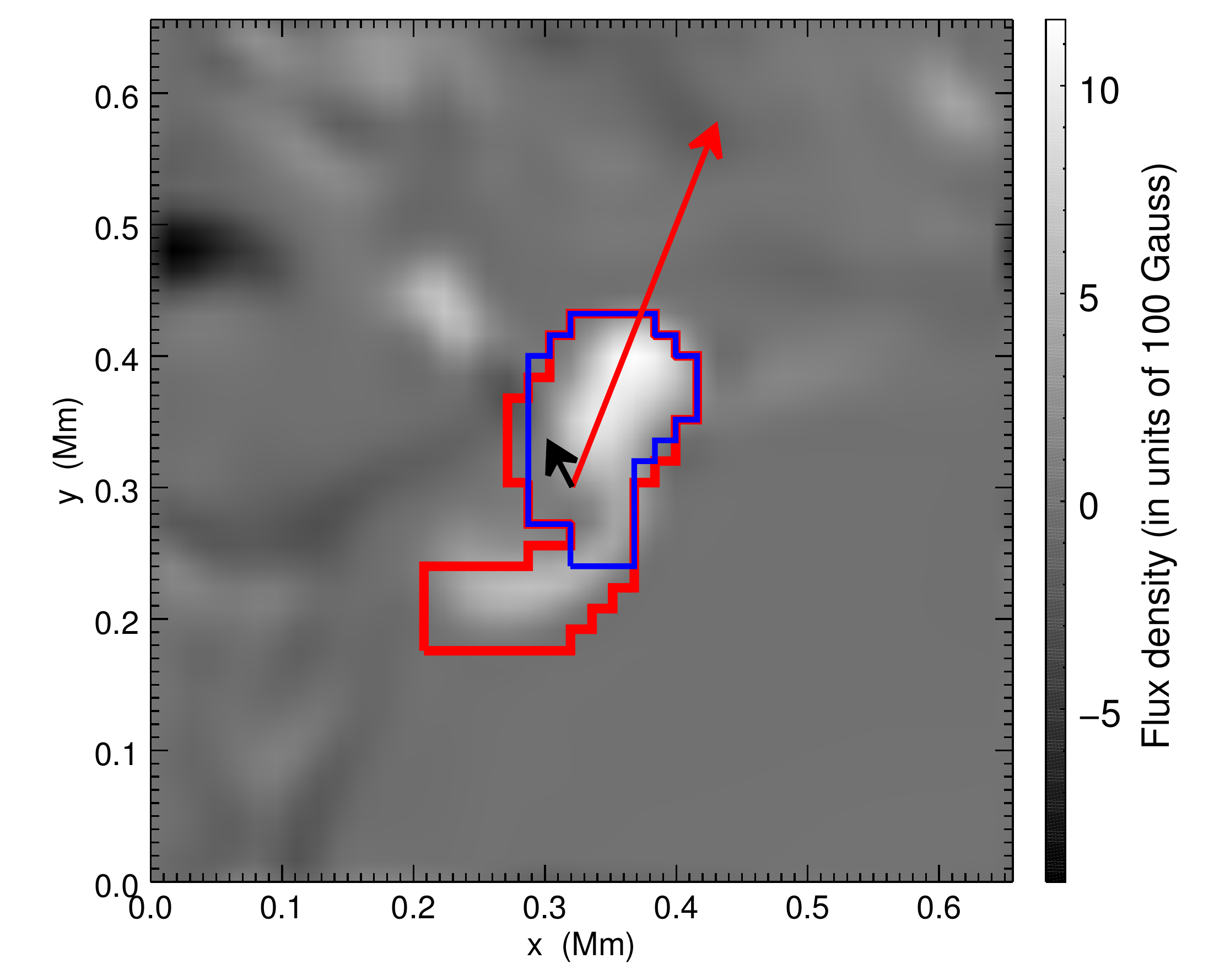} 
	\caption{\textit{Red} and \textit{blue contours} mark the boundary of the area covered by a flux element at two consecutive time-steps (\textit{blue} being the later time). \textit{Red} and \textit{black arrows} are scaled with respect to each other and signify the differing magnitude and direction of the displacement of magnetic element barycenter due to flux evolution and plasma flow, respectively.}
	\label{fig-muram-visual}
\end{figure}

We suggest that both the non-ballistic scaling of the element displacements at short time intervals and spurious supersonic Lagrangian velocities deduced from the element motions in the simulations are a consequence of using magnetic elements' barycenters in determining their positions and misinterpreting all changes in those positions as true motions.  Changes in the magnetic flux content of an element subjects its barycenter to random changes in position. This introduces jitter that is erroneously interpreted as motion. Figure~\ref{fig-muram-visual} presents an illustrative example.  The outline of the magnetic element at two consecutive time-steps is shown in \textit{red} and \textit{blue}, with \textit{blue} being the later time.  The \textit{red arrow} indicates the displacement (scaled for illustration), as computed from the shift in element's barycenter.  As the magnetic element evolves, the disappearance of flux displaces its barycenter with a magnitude and direction drastically different from what would be achieved by passive advection due to underlying plasma flow.  The \textit{black arrow} in the figure (scaled as the {\it red arrow})  indicates the displacement computed using the vector average of the plasma velocities over all grid-points constituting the element at the initial time-step.  Magnetic flux evolution thus introduces a random component to the deduced motions of flux element barycenters which dominates the scaling at shortest time intervals and can lead to unrealistically large Lagrangian velocity values.  This is true irrespective of whether the barycenter definition is flux-weighted or position-weighted.

To further assess the impact of this jitter, we compared magnetic element motions with passive tracer motions in the simulation.  The motions of passive tracers were evolved using the photospheric horizontal plasma velocity at their locations, and are independent of any contribution from flux evolution.  For point-like passive tracers seeded at random locations, the scaling for the shortest temporal increments is ballistic.  To ensure that the flux element positions are not biased, we separately seeded the passive tracers co-spatially with the flux elements, so that they initially occupied the same area (the same grid-point locations) as the corresponding elements. The passive tracer `element' positions were then evolved using the horizontal velocity field averaged over the area they spanned. This traces the motions of their position-weighted barycenter and allows direct comparison between the flux element and passive tracer statistics if the field distribution over the elements is nearly uniform.  This approximation holds for the magnetic elements identified using the employed three standard-deviation threshold, and for these elements we found consistent velocity and displacement statistics independent of the barycenter definition.  

The barycenter velocity distribution of the passive tracer `elements' shows no pathologically high values (\textit{black} curves in Figure~\ref{fig-muram-pdf-vel}), and for short temporal increments, the mean squared displacement of the `elements' scales ballistically (for \textit{black dots} in Figure~\ref{fig-muram-drdt}), in contrast to the super-diffusive scaling for magnetic elements (for \textit{red stars} in Figure~\ref{fig-muram-drdt}).  For longer temporal increments, the mean squared displacement scaling of the flux elements and passive tracers agree, suggesting that over longer timescales the passive advection of flux elements due to the plasma flow dominates the random barycenter jitter.  For sufficiently long increments (not shown in the plot), and in the absence of any large-scale flow (discussed in~\S\ref{subsec-model-dist}), the slope for both curves should approach the diffusive value, though direct comparison between the flux element and passive tracer statistics makes the most sense over short temporal increments over which the passive tracer and element locations remain contiguous.

The scaling behavior of the magnetic flux elements in  \muram\ (for {\it red stars} in Figure~\ref{fig-muram-drdt}) can be reproduced using a simple two-component model of their motion:  motion due to a correlated random walk (corresponding to passive advection by granular flows) and motion due to delta-correlated jitter (mimicking random barycenter motion due to flux evolution). The correlated motion follows the same formulation as that in \S\ref{sec-model-const} without the supergranular drift component ($C=8.4$, $v_{\rm g}=$ 2.0 km/s, $v_{\rm sg}=0$ km/s). The parameter $C$ is chosen so that, when scaled with \muram\ time-step $\Delta t=2.0625$ seconds, the motion de-correlates over a typical granular lifetime (about 6 minutes).  The granular velocity $v_{\rm g}$ is chosen to be the mean of the passive tracer barycenter velocity magnitudes when the tracers are seeded co-spatially with the flux element locations (as described above). 
We note that $v_{\rm g}$ here is distinct from that used in~\S\ref{subsec-model-dist}.  Here, since it is based on passive tracers in the plasma velocity field, it does not include the barycenter jitter due to flux element evolution.  The jitter component is instead explicitly modeled as a vector of constant magnitude with random orientation and is added to the correlated granular motions at each time-step.  Plotted in Figure~\ref{fig-muram-drdt} with \textit{blue squares} is the model displacement when using the values above, along with a jitter magnitude 3.5 km/s.  The value of 3.5 km/s is arbitrary but demonstrates that this model can capture the scaling observed for the magnetic flux elements over short temporal increments.  The real distribution of jitter velocity magnitudes in the simulation can be determined by the difference between the measured flux element barycenter values ({\it red} distributions in Figure~\ref{fig-muram-pdf-vel})  and their passive tracer counterparts  ({\it black} distributions in Figure~\ref{fig-muram-pdf-vel}).  It is broad with a mean value of 2.3 km/s and a standard-deviation of 3.8 km/s.  What is important is that this highly simplified model captures the effect of both the correlated granular flow (which would alone yield ballistic scaling) and the random apparent motions due to flux element evolution (which would alone yield diffusive scaling). Together these two components explain the super-diffusive scaling at the shortest time intervals and the increase in scaling exponent as the temporal increment increases.  

\subsection{Magnetic element displacement vs. image resolution}
\label{additionalstudy}

\begin{table}[t]
	\centering
	\caption{Scaling vs. image resolution, computed for flux elements tracked in \muram\ simulations}  
	\label{table-scaling-vs-resolution}
	\begin{tabular}{lllll} 
		\hline
		Pixel resolution    & Flux density threshold$^a$  &  ${\gamma_0}^b$ 	 	& ${\sigma_v}^c$ \\ 
		(km)		 		 & (Gauss)  	      &     		     		& (km/s)  \\      
		\hline
		16 			  & 470 				& 1.21 		 		  			& 3.39	\\		
		27 			  & 412 		        & 1.32 	    		  			& 2.74 	\\		%
		38.5 		  & 377 		        & 1.44 		 		  			& 2.43	\\		
		47 		      & 354 			  	& 1.50 		 		  			& 2.25	\\		
		100		      & 254 				& 1.73 				  			& 1.62 	\\		
		116		      & 234 				& 1.75 	      		  			& 1.58	\\		
		\hline
	\end{tabular}
	\\
	\raggedright{$^a$average of three times the standard-deviation values for all magnetograms, $^b$scaling at the shortest temporal increments, $^c$standard-deviation of the Lagrangian velocity distribution of the flux elements.} 
	\\
\end{table}

The barycenter jitter contribution to magnetic element motion is sensitive to the physical processes governing flux evolution, image resolution and cadence, and the feature tracking algorithm and identification parameters.  \textit{Green diamonds} in Figure~\ref{fig-muram-drdt} illustrate the effect of image resolution on the mean squared displacement.  Flux elements were identified and tracked after convolving the \muram\ images with a two-dimensional Gaussian kernel, degrading the \muram\ image resolution to $\sim$116 km (Gaussian full-width at half maximum equal to the ratio of target pixel size and \muram\ pixel size). The scaling exponent obtained (for \textit{green diamonds}) has a larger value at the shortest increments than that obtained when tracking elements at the native resolution (for \textit{red stars}).  Table~\ref{table-scaling-vs-resolution} shows the monotonic increase in the short increment scaling exponent $\gamma_0$ with decreasing resolution.  While at these timescales the flow contribution to the displacement is ballistic, the barycenter jitter contribution to the scaling decreases with decreasing resolution. This is because the flux elements being tracked in a degraded image are on average larger so their barycenter positions are less sensitive to small changes in the field configuration, the probability for large changes in flux density is reduced by the reduction in resolution, and the shape of the elements in a low-resolution image are less irregular making the barycenter definition more robust and its position less sensitive to barycenter jitter.     

\edited{It is important to also note that the plot for the low-resolution results for flux elements (\textit{green diamonds}) in Figure~\ref{fig-muram-drdt} were shifted vertically upward to better compare its shape with the native resolution (\textit{red star}) displacements at longest increments.} The magnitude of the displacement depends on the Lagrangian velocity distribution which decreases in width with decreasing resolution ($\sigma_v$ in Table~\ref{table-scaling-vs-resolution}). The velocity distribution captures both the advective and the jitter contributions, both of which are reduced when the image is degraded. This leads to smaller mean squared displacements for the same temporal increment. Thus, both the scaling coefficient at the shortest time intervals and the magnitude of the mean squared displacements are not robust and depend on the image properties and the feature tracking parameters.

\section{Summary and Conclusion}
\label{sec-summary}

We examined the transport of internetwork magnetic flux elements in \textit{Hinode}/NFI data and found, as in previous studies~\citep[e.g.,][]{chitta12, abramenko11, jafarzadeh14}, that their mean squared displacement scales super-diffusively with time even for the shortest temporal increments. In addition, the shape of the underlying displacement probability distribution evolves from Rayleigh to Gaussian as the increment increases.  Using a correlated random walk model with a drift component, we have demonstrated that this is likely due to supergranular motions dominating granular motions for time intervals long compared to the granular correlation time.  We suggest that over intervals longer than supergranular correlation times, the distributions would be similarly affected by the underlying meridional flow. Thus, the interpretation of flux element motion as a strictly diffusive process is likely incorrect. 

Super-diffusive scaling is found in studies using observational data with cadences as short as 5 seconds~\citep[e.g., ][]{chitta12}, much shorter than the expected Lagrangian and Eulerian integral times of the flow and thus capturing timescales over which the displacement scaling should be ballistic.  We investigated the underlying causes for this discrepancy by tracking flux elements in a \muram\ simulation with 2 second cadence and found similar super-diffusive scaling for short temporal increments.  Comparison between the flux element and passive tracer statistics in the simulation suggests that the super-diffusive scaling over short temporal intervals is a consequence of misinterpreting flux element barycenter motion as strictly due to plasma flows.  In addition to motions induced by the underlying flow, barycenter positions are subject to jitter induced by magnetic flux evolution.  This imparts a random component to the measured motions and contributes strongly to the scaling at the shortest temporal increments and results in the observed super-diffusive scaling.  The measured Lagrangian velocity distribution reflects these spurious motions as well, showing values well in excess of the photospheric sound speed.  Moreover, the jitter contribution depends on the underlying physical processes governing flux evolution, image resolution and cadence, and the magnetic element identification scheme employed. 

These results suggest that using displacement measurements of flux element barycenters to directly determine diffusion coefficients or wave forcing by magnetic element motions in the solar photosphere may be problematic. However, the artifacts identified may be partially overcome by employing passive tracers as a proxy, as their motion is independent of flux evolution. 
Rather than tracking magnetic elements over long periods of time, it may be preferable to compute the photospheric horizontal plasma velocity using methods such as structure or correlation tracking\edited{~\citep[e.g.,][]{Simon88, Roudier99, Potts04, Roudier12, Attie16}}. The velocity field could then be used to compute passive tracer trajectories and displacement statistics.  
In addition to likely being more robust, this would avoid the ubiquitous difficulties associated with flux element sparsity, identification and evolution, and allow the investigation of displacement along many more trajectories than is usually possible, with each lasting longer than typical magnetic element lifetimes.    
From the trajectories, one could measure both the drift contribution of larger scale flows and the effective diffusive component of the random motions by fitting the observed displacement probability distribution with the analytic function of Equation~\ref{eqn-pdf1}.  The fit would yield both the width of the displacement distribution $\sigma(t)$, a measure of the diffusive component of the motion, and the average displacement due to the drift motion $r_0(t)$.  The success of this method would rely on the accuracy of the derived photospheric plasma velocity field and the ability to separate  the diffusive and the drift components if the large-scale flow is not uniform in space and time, but from this work, we expect the method to yield a more authentic measure of the magnetic flux element motion.
\\

This paper is based on the data acquired during Hinode Operation Plan 151. We thank the Hinode Chief Observers for their efforts in executing this plan.  Hinode was developed and launched by ISAS/JAXA with NAOJ as a domestic partner and NASA and STFC (UK) as international partners. It is operated by these agencies in cooperation with ESA and NSC (Norway). This work has been partially funded by the Spanish Ministerio de Econom\'{\i}a y Competitividad through projects ESP2013-47349-C6-1-R and ESP2016-77548-C5-1-R including European FEDER funds. The research has made use of NASA's Astrophysics Data System Bibliographic Services. 
NCAR is supported by the National Science Foundation. 
The authors thank Samuel Van Kooten for magnetic bright points tracking. MPR was partially supported by NASA award NNX12AB35G. 
P. Agrawal acknowledges the support of the University of Colorado's George Ellery Hale Graduate Student Fellowship. 


\appendix
\section{Bivariate transformation of random variables}
\label{appendixa}
To determine the probability density of a function $u=f(x,y)$ of independent random variables $x$ and $y$, with individual probability densities $P(x)$ and $P(y)$ and joint probability density $P_{xy}(x,y)=P(x)P(y)$, define a second function $v=g(x,y)$ chosen so that its probability density can be integrated out of the joint probability density, leaving that of the target function behind.  For $u$ and $v$ with inverses $x=h_1(u,v)$ and $y=h_2(u,v)$, the joint probability density of $u$ and $v$ is~\citep[e.g.,][]{casella2002statistical, hogg2006probability}
\begin{equation}
\label{eqna1}
P_{uv}(u,v)=P_{xy}(h_1,h_2)
\left\vert{\begin{array}{cc}
{{\partial h_1}\over{\partial u}}&{{\partial h_1}\over{\partial v}}\\
{\ }&{\ }\\
{{\partial h_2}\over{\partial u}}&{{\partial h_2}\over{\partial v}}
\end{array}}\right\vert
\end{equation}
and 
\begin{equation}
P(u)=\int\!\!P(u,v)\,dv \ .
\end{equation}

To derive Equation~\ref{eqn-pdf1} in Section~\ref{subsec-model-dist}, consider $P(r)$ with $r=\sqrt{x^2+y^2}$, where as in~\S\ref{subsec-model-dist}, $x$ and $y$ are independent random spatial offsets Gaussianly distributed about $x_0$ and $y_0$ with equal variance $\sigma$. By bivariate transformation, 
the joint probability density
\begin{equation}
P_{xy}(x,y)=\frac{1}{2\pi\sigma^2}e^{-\left[(x-x_0)^2+(y-y_0)^2\right]/2\sigma^2}\ .
\end{equation}
can be written in terms of $r$ ($u$ in the general notation above) and $\theta=\tan^{-1}(y/x)$ ($v$ in the general notation above), 
with inverse functions $x=r\cos\theta$ ($h_1$ in the general notation above) and $y=r\sin\theta$ ($h_2$ in the general notation above), by evaluating Equation~\ref{eqna1},
\begin{equation}
P_{r\theta}(r, \theta)= \frac{r}{2\pi\sigma^2}e^{-\left[r^2 + r_0^2 -2r_0r\cos(\theta-\phi)\right]/2\sigma^2}=\frac{r}{2\pi\sigma^2} \ e^{-(r-r_0)^2/2\sigma^2}\ e^{-r_0r/\sigma^2}\ e^{r_0rcos(\theta-\phi)/\sigma^2}\ ,
\end{equation}
with $\phi = \tan^{-1}(y_0/x_0)$ and $r_0=\sqrt{x_0^2+y_0^2}$.  Integrating over all $\theta$ then yields the probability density of $r$ (Equation~\ref{eqn-pdf1} in the main text),
\begin{equation}
P(r) = \frac{r}{2\pi\sigma^2} \ e^{-(r-r_0)^2/2\sigma^2}\ e^{-r_0r/\sigma^2}\ \int_{0}^{2\pi} e^{r_0rcos(\theta-\phi)/\sigma^2}\ d\theta = \frac{r}{\sigma^2} \ e^{-(r-r_0)^2/2\sigma^2}\ e^{-r_0r/\sigma^2}\ I_0(r_0r/\sigma^2)\ ,
\end{equation}
where, $I_0$ is the lowest order modified Bessel function of the first kind~\citep{book-abramowitz72}.

\end{document}